\documentclass{article} 
\usepackage{iclr2026_conference,times}
\usepackage{graphicx}
\usepackage{booktabs}
\usepackage{hyperref}
\usepackage{url}
\usepackage{xspace}
\usepackage{enumitem}
\usepackage[capitalize]{cleveref}
\crefname{section}{Sec.}{Secs.}
\crefname{section}{Section}{Sections}
\crefname{table}{Table}{Tables}
\crefname{table}{Tab.}{Tabs.}

\title{\name: Execution-Grounded Multi-Agent Test Oracle Synthesis}

\author{Dong HUANG, Mingzhe Du, Jie M. Zhang, Zheng Lin, Meng Luo, Regina Zhang, See-Kiong Ng
\\
\texttt{\{dhuang,mingzhe\}@nus.edu.sg} \\
}

\newcommand{\name}{Nexus\xspace}

\iclrfinalcopy 
\begin{document}

\maketitle

\begin{abstract}
Test oracle generation in non-regression testing is a longstanding challenge in software engineering, where the goal is to produce oracles that can accurately determine whether a function under test (FUT) behaves as intended for a given input.
In this paper, we introduce \name, a novel multi-agent framework to address this challenge. \name generates test oracles by leveraging a diverse set of specialized agents that synthesizes test oracles through a structured process of deliberation, validation, and iterative self-refinement.
During the deliberation phase, a panel of four specialist agents, each embodying a distinct testing philosophy, collaboratively critiques and refines an initial set of test oracles.
Then, in the validation phase, \name generates a plausible candidate implementation of the FUT and executes the proposed oracles against it in a secure sandbox.
For any oracle that fails this execution-based check, \name activates an automated self-refinement loop, using the specific runtime error to debug and correct the oracle before re-validation.
Our extensive evaluation on seven diverse benchmarks demonstrates that \name consistently and substantially outperforms state-of-the-art baselines.
For instance, \name improves the test-level oracle accuracy on the LiveCodeBench from 46.30\% to 57.73\% for GPT-4.1-Mini. 
The improved accuracy also significantly enhances downstream tasks: the bug detection rate of GPT-4.1-Mini generated test oracles on HumanEval increases from 90.91\% to 95.45\% for \name compared to baselines, and the success rate of automated program repair improves from 35.23\% to 69.32\%.
\end{abstract}

\section{Introduction}

Unit testing represents a cornerstone of modern software engineering, enabling developers to verify the correctness of individual code components in isolation~\citep{beck2022test}.
A well-crafted unit test comprises two essential components: a \textit{test input} that exercises a specific execution path within the function under test (FUT), and a \textit{test oracle} that determines whether the function's actual output aligns with its expected behavior~\citep{hossain2024togll}.
Despite their critical importance, constructing comprehensive unit tests remains a labor-intensive endeavor that demands not only deep domain expertise but also meticulous attention to edge cases and complex interaction patterns~\citep{daka2015modeling,6982627}.

To reduce the burden of manual test construction, researchers have developed a wide range of automated test generation techniques, from fuzzing~\citep{fioraldi2023dissecting,miller1990empirical}, symbolic execution~\citep{cadar2008exe,godefroid2005dart}, and search-based testing~\citep{fraser2011evolutionary,fraser2011evosuite} to modern LLM-based approaches~\citep{du2024evaluating,qing2025effibench,huang2024effibench}. However, these methods largely prioritize generating diverse test inputs for code coverage, while oracle generation remains secondary due to the long-standing \emph{oracle problem}~\citep{barr2014oracle}—the difficulty of determining correct outputs for non-trivial programs. As a result, most work rely on \textit{regression oracles}, which record outputs from the current implementation as expected behavior. 
While valuable for detecting behavioral changes, regression oracles fundamentally cannot verify the functional correctness of new or modified code against its intended specification.


The challenge of generating specification-based oracles that verify correctness against intended behavior rather than existing implementation has emerged as a critical bottleneck in automated testing.
Moreover, it brings significant obstacles to LLM-based and agent-based automatic code generation, where reliable feedback is essential for enabling LLMs to iteratively refine their outputs.

Recent advances in LLMs have opened new avenues for addressing this challenge, with CANDOR~\citep{xu2025hallucination} representing the state-of-the-art, introducing a multi-agent framework where multiple LLM agents collaborate through panel discussions to reach consensus on oracle predictions.
However, our analysis reveals two key limitations of CANDOR: single-paradigm reasoning and a deliberation-only architecture.
First, all agents rely on the same source of truth (i.e., the natural language documentation of the FUT), and play similar roles based on identical instructions. This creates an epistemological bottleneck, as the agents share the same interpretation of the specification.
Second, reasoning is confined to abstract, language-based deliberation without grounding in execution. Since proposed oracles are never validated against even a plausible implementation, the framework lacks feedback to detect subtle bugs or undocumented edge cases.


\begin{figure}
    \centering
    \includegraphics[width=\linewidth]{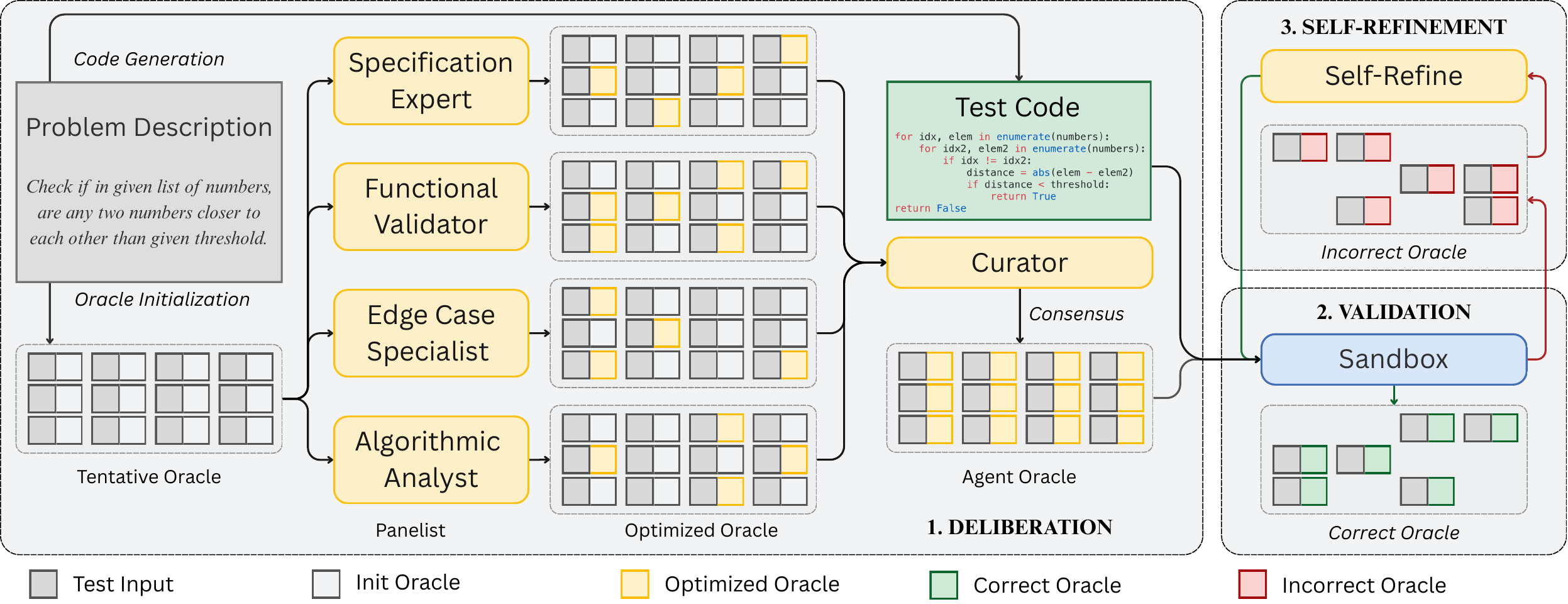}
    \vspace{-1.8em}
    \caption{The architecture of the \name framework. 
        \textbf{(1) Deliberation:} A panel of specialist agents collaborates to produce candidate oracles. 
        \textbf{(2) Validation:} Oracles are validated by executing them against a plausible, LLM-generated implementation of the function under test. 
        \textbf{(3) Self-Refinement:} Failed oracles enter an iterative loop where runtime errors are used for automated debugging. This pipeline grounds abstract deliberation in executable evidence.}
    \label{fig:framework}
\end{figure}

In this paper, we introduce \textbf{\name}, a novel multi-agent framework that overcomes this limitation by grounding language-based reasoning in executable evidence.
First, to break the epistemological echo chamber, the framework begins with a deliberation phase powered by a panel of four specialist agents, each embodying a distinct and orthogonal testing philosophy: a \emph{Specification Expert} for literal compliance, an \emph{Edge Case Specialist} for adversarial robustness, a \emph{Functional Validator} for high-level semantic properties, and an \emph{Algorithmic Analyst} for implementation-level logic.
This multi-perspective analysis ensures a far more comprehensive initial critique than a panel merely re-examining the same specification.
Second, to transcend the limitations of any purely text-based analysis, \name introduces a validation phase.
Here, the framework generates a plausible candidate implementation of the FUT and executes the proposed oracles against this code in a secure sandbox, providing a powerful, execution-based source of truth that is independent of the original documentation.
Finally, to capitalize on this new source of feedback, \name activates a self-refinement loop for any failed oracle.
This loop uses the specific runtime error from the sandbox to provide a precise and actionable signal to automatically debug and correct the oracle before re-validation.

Our extensive evaluation on seven diverse benchmarks demonstrates that \name consistently and substantially outperforms state-of-the-art baselines across a suite of open-source and proprietary models.
The superiority of \name is not marginal; for instance, when powered by GPT-4.1-Mini, \name boosts test-level oracle accuracy on the challenging LiveCodeBench benchmark to $57.73\%$ from the state-of-the-art's $46.30\%$ achieved by CANDOR.
The improved accuracy also significantly enhances downstream tasks: the bug detection rate of GPT-4.1-Mini generated test oracles on HumanEval increases from 90.91\% to 95.45\% for \name compared to baselines, and the success rate of automated program repair improves from 35.23\% to 69.32\%.
By uniquely combining multi-agent deliberation with validation and self-refinement, \name establishes a new paradigm for generating reliable, specification-grounded test oracles. The main contributions of this work are as follows:

\begin{itemize}[leftmargin=*]
    \item We design and implement \name, a novel multi-agent framework that synthesizes test oracles through a unique pipeline of deliberation, validation, and self-refinement, overcoming the single-source-of-truth limitation in prior work.
    \item We introduce a panel of four specialist agents, each embodying a distinct testing philosophy, to facilitate a more comprehensive and effective deliberation process for oracle generation.
    \item We conduct a comprehensive evaluation on seven benchmarks using four different LLMs, demonstrating that \name significantly outperforms state-of-the-art methods in oracle accuracy and enhances downstream tasks like bug detection and automated program repair.
\end{itemize}

\section{Related Work}

The generation of test oracles presents fundamentally different challenges from input generation, requiring a deep understanding of intended program behavior rather than mere code coverage.
Early automated testing tools largely sidestepped this challenge by employing regression oracles or simple crash oracles~\citep{fraser2011evosuite,pacheco2007randoop}.
While regression oracles effectively detect behavioral changes between versions, they cannot validate functional correctness against specifications, potentially perpetuating bugs across software evolution~\citep{pizzorno2024coverup}.
The specification oracle problem that generating oracles that verify correctness against intended behavior has long been recognized as one of the most challenging aspects of automated testing~\citep{barr2014oracle}. The emergence of LLM-based approaches brought new possibilities for oracle generation.
TOGA~\citep{toga2022} pioneered the use of transformer models for inferring specification-based oracles, fine-tuning CodeBERT to predict expected outputs from code context.
This work inspired subsequent research exploring various neural architectures and training strategies for oracle generation~\citep{endres2024can,hayet2024chatassert}.
TOGLL~\citep{hossain2024togll} achieved state-of-the-art performance by combining EvoSuite-generated inputs with fine-tuned LLM oracles, demonstrating that hybrid approaches could leverage the strengths of both traditional and learning-based methods. However, fine-tuning approaches face significant limitations in generalization, as model performance degrades substantially when test distributions shift from training data~\citep{xu2025hallucination}.

Recent work has increasingly focused on prompt engineering strategies that leverage the zero-shot capabilities of LLMs.
\citet{siddiq2024using} demonstrated that carefully crafted prompts could enable LLMs to generate both inputs and oracles from natural language descriptions without fine-tuning.
RetriGen~\citep{zhang2025improving} enhanced oracle generation through retrieval-augmented generation, incorporating relevant assertions from external codebases to guide prediction.
CANDOR~\citep{xu2025hallucination} represents the current SOTA method, which introduces a multi-agent framework where several LLM agents collaborate through a panel discussion to reach a consensus on the correct oracle.
While CANDOR achieves impressive results without fine-tuning, our analysis reveals that its single-paradigm and deliberation-only approach limits its ability to detect bugs beyond explicit specifications.

In this work, we propose \name, which overcomes these limitations by introducing our novel deliberation-validation-refinement architecture.
\name employs a multi-agent framework that facilitates diverse reasoning strategies, enhancing the initial analysis of test oracles.
By incorporating a validation phase that grounds oracles in executable contexts and a self-refinement loop that leverages runtime feedback, \name further improves the accuracy of generated oracles.

\section{Methodology}


\subsection{Preliminary Analysis}
To empirically validate our assertion that oracle generation is the predominant challenge in automated test generation for LLMs, we conducted a preliminary case study.
In this study, we prompted GPT-4.1-nano to generate complete unit test cases from the natural language descriptions provided in a diverse set of standard benchmarks.
We then performed a manual root cause analysis for each incorrectly generated test case.
Each failure was categorized based on its origin: errors stemming from an invalid or logically flawed test input, versus those caused by an incorrect test oracle. The evaluation results are shown in \cref{tab:study1}, demonstrating a clear and consistent trend across all seven benchmarks, i.e., the majority of errors are attributable to incorrect test oracles.
For instance, on the HumanEval benchmark, $88.89\%$ of errors were due to an incorrect oracle, compared to only $11.11\%$ from invalid test inputs.
This finding strongly suggests that while LLMs can proficiently generate syntactically valid and plausible inputs, they struggle profoundly with the deep semantic reasoning required to predict the correct output.
\begin{table}
    \centering
    \caption{Error distribution of the incorrect test cases.}
    \label{tab:study1}
    
    \resizebox{\textwidth}{!}{
    \begin{tabular}{l|ccccccc}
    \toprule
         Error& HumanEval & MBPP & HumanEval-Pro & MBPP-Pro & TestEval & LiveCodeBench&ULT\\
         \midrule
         Test Input&11.11&15.38&8.57&17.14&0.00&20.29&9.38\\
         Test Output&88.89&84.62&91.43&82.86&100.00&79.71&90.62\\

         \bottomrule
    \end{tabular}
    }
\end{table}

\subsection{Framework Overview}
To address this critical oracle problem, we introduce \name, a multi-agent framework that synthesizes test oracles through a structured process of deliberation, validation, and iterative self-refinement. As shown in \cref{fig:framework}, unlike prior works that rely on a single source of truth, \name orchestrates a comprehensive pipeline where candidate oracles are first debated by a panel of specialist agents, then consolidated by a curator, and finally subjected to rigorous execution-based testing.
The core novelty of our framework lies in its final two stages: a validation phase that checks oracle correctness against a plausible, LLM-generated implementation of the FUT, and a self-refinement loop that uses execution feedback to automatically debug and correct any oracles that fail this check.
This architecture grounds the abstract reasoning of LLMs in concrete, executable evidence, creating an effective feedback mechanism that significantly enhances test oracle accuracy.

\subsection{Multi-Phase Architecture}
The \name framework operates as a structured, three-phase pipeline: multi-agent deliberation, execution-based validation, and iterative self-refinement. Each phase progressively scrutinizes and improves a set of candidate oracles, culminating in a final set of validated and self-corrected test oracles.

\noindent\textbf{Phase 1: Multi-Agent Deliberation.}
The process begins by generating a set of \textit{Tentative Oracles} using a standard zero-shot prompting approach. These initial oracles serve as concrete hypotheses for a structured, hierarchical debate among a panel of four specialist agents, each embodying a specific testing persona to ensure diverse and comprehensive analysis.
The \textbf{Specification Expert} focuses on adherence to documented specifications and requirements, ensuring strict fidelity to the function's explicit contract. 
It systematically cross-references each candidate oracle against the formal requirements extracted in Phase 1, verifying adherence to documented constraints such as return types, value ranges, and precisely defined behaviors.
The \textbf{Edge Case Specialist} focuses on boundary conditions, corner cases, and error scenarios.
It probes the oracle's validity under boundary conditions and with atypical inputs often overlooked in documentation, such as empty collections, null values, zeros, and data type extremities.
The \textbf{Functional Validator} focuses on core functionality and expected input-output relationships.
It concentrates on the core purpose of the function, ensuring the input-output relationship is logically sound and aligns with general programming principles, even for behaviors not explicitly detailed.
For instance, for a sort function, it would verify that the output is a permutation of the input, a fundamental property that may not be explicitly stated in the documentation.
The \textbf{Algorithmic Analyst} predicts the correct output by performing a conceptual execution of a plausible algorithm.
Based on the function's description, it infers a likely implementation strategy. It performs a step-by-step mental trace for a given test input to compute the expected result, making it effective at catching subtle off-by-one or logic errors.
A \textit{Curator} agent then synthesizes these multiple perspectives, resolves disagreements, and produces a single, consolidated set of \textit{Candidate Oracles} that represents the collective judgment of the deliberation process.

\noindent\textbf{Phase 2: Execution-Based Validation}
To ground the abstract deliberation in concrete evidence, this phase tests the candidate oracles against a plausible implementation. First, an LLM is prompted to generate a \textit{Candidate Implementation} of the FUT based on its documentation. This generated code serves as a reasonable, albeit not guaranteed to be perfect, model of the intended behavior. The candidate oracles are then executed as test assertions against this implementation in a secure \textit{Code Execution Sandbox}. This step provides empirical feedback: oracles that pass are considered highly likely to be correct, while those that fail are flagged as erroneous and passed to the next phase.

\noindent\textbf{Phase 3: Iterative Self-Refinement}
For every oracle that fails validation, an automated refinement loop is activated. The framework constructs a detailed debugging prompt containing the FUT specification, the candidate implementation, the failed oracle, and the specific error message from the sandbox. An LLM is then prompted to generate a corrected oracle. This refinement process is iterative: the proposed fix is sent back to the \textit{Code Execution Sandbox} for re-validation. This loop continues for a predefined number of iterations or until the oracle passes, ensuring the final oracles have successfully withstood rigorous, execution-based scrutiny.

\section{Evaluation}
\label{sec:evaluation}


Our evaluation is structured to address three key research questions:
\begin{itemize}
    \item \textbf{RQ1} investigates how \name performs in terms of oracle accuracy compared to state-of-the-art baselines across different models and benchmarks.
    \item \textbf{RQ2} analyzes the contributions of the core components of the \name framework to its overall performance.
    \item \textbf{RQ3} examines how the number of iterative refinement steps in the self-correction loop impacts the final oracle accuracy.
\end{itemize}

\subsection{Experimental Setup}

\paragraph{Task Definition}
We focus on the task of test oracle generation.
In this task, the model is provided with the natural language description of a function (e.g., its docstring) and a predefined test input.
The goal is for the model to generate the corresponding test oracle, which is the expected output for that given input, typically formatted as a programmatic assertion.
To simulate a fully automated testing pipeline, we use test inputs that are themselves generated by GPT-4.1-nano\footnote{For each programming task, we generate 20 diverse test inputs and keep them fixed across all experiments to ensure fair and consistent comparisons. These inputs are available in our replication package.}.

\paragraph{Datasets}
To ensure our findings are generalizable, we conduct experiments across seven diverse and widely-used benchmarks for code intelligence.
HumanEval~\citep{ChenCodex2021} and MBPP~\citep{Austin2021} are canonical benchmarks for evaluating code generation, containing a range of programming problems.
We also include their more challenging counterparts, HumanEval-Pro and MBPP-Pro~\citep{yu2024humanevalpro}, which feature more complex requirements and edge cases.
Furthermore, we incorporate LiveCodeBench~\citep{jain2024livecodebench}, a benchmark designed to be robust against data contamination from web-scale training sets.
Finally, to specifically assess performance on testing-related tasks, we include TestEval~\citep{wang2025testeval} and ULT~\citep{huang2025benchmarking}, which are benchmarks created explicitly for evaluating test generation capabilities.

\paragraph{Models}
To demonstrate the broad applicability of our framework, we evaluate its performance on a suite of both open-source and closed-source LLMs, representing a range of sizes and capabilities.
Our selection of open-source models includes DeepSeek-R1-Distill-Qwen-7B (\textbf{Qwen-7B}) and DeepSeek-R1-Distill-Qwen-32B (\textbf{Qwen-32B}).
For closed-source models, we use OpenAI's \textbf{GPT-4.1-nano} and \textbf{GPT-4.1-mini}.
For all experiments, we set the decoding temperature parameter to $0.0$ to ensure deterministic and reproducible outputs.

\paragraph{Baselines}
We compare \name against two primary baselines to contextualize its performance.
The first is \textbf{Direct Generation}, which represents the standard, zero-shot prompting approach where the LLM is instructed to generate the test oracle without any advanced multi-agent framework.
This reflects the standard practice for naive test generation.
The second is \textbf{CANDOR}~\citep{xu2025hallucination}, the current state-of-the-art multi-agent framework for test oracle generation.
As detailed in our introduction, CANDOR relies on a consensus-based panel discussion where all agents derive their reasoning from the same specification source, making it a strong, contemporary baseline.

\paragraph{Evaluation Metrics}
To quantitatively measure performance, we define three key metrics that assess both the intrinsic correctness of the oracles and their extrinsic value in downstream tasks.
Our primary metric is \textbf{Oracle Accuracy}, which measures the percentage of generated test oracles that are semantically correct, as determined by execution against the canonical benchmark solution.
To assess practical utility, we also measure the \textbf{Bug Detection Rate}, defined as the percentage of buggy programs from a curated set that are successfully identified by our generated test cases.
Finally, we evaluate the downstream impact on automated program repair with \textbf{Self-Debugging Performance}, measured by the final Pass@1 rate of repaired code against the official hidden test suite after receiving feedback from a generated oracle.

\begin{table}
\centering
\caption{Accuracy of LLM-generated test oracles at both the task (Pass@1) and test-case levels. The value in parentheses denotes the absolute improvement over the Direct Generation baseline. \textbf{HE}: HumanEval, \textbf{HE-Pro}: HumanEval-Pro, \textbf{TE}: TestEval, \textbf{LCB}: LiveCodeBench.}
\label{tab:main_accuracy}
\resizebox{\textwidth}{!}{%
\begin{tabular}{l|ccccccc}
\toprule
Method & HE & HE-Pro & MBPP & MBPP-Pro & TE & LCB & ULT \\
\midrule
\multicolumn{8}{c}{\textbf{Task Level Accuracy (\%)}} \\
\midrule
Qwen-7B & 10.98 & 1.22 & 14.40 & 3.70 & 0.00 & 0.93 & 0.00 \\
+ CANDOR & 6.10 (-4.88) & 1.83 (+0.61) & 14.79 (+0.39) & 2.65 (-1.06) & 0.49 (+0.49) & 0.00 (-0.93) & 0.00 (+0.00) \\
+ \name & \textbf{13.41 (+2.43)} & \textbf{3.66 (+2.44)} & \textbf{20.23 (+5.83)} & \textbf{5.03 (+1.33)} & \textbf{0.49 (+0.49)} & \textbf{3.74 (+2.81)} & \textbf{0.85 (+0.85)} \\
\midrule
Qwen-32B & 12.80 & 4.27 & 24.51 & 9.79 & 0.00 & 0.93 & 1.71 \\
+ CANDOR & 6.10 (-6.70) & 3.66 (-0.61) & 7.39 (-17.12) & 2.65 (-7.14) & 0.00 (+0.00) & 8.41 (+7.48) & 0.85 (-0.86) \\
+ \name & \textbf{29.88 (+17.08)} & \textbf{10.37 (+6.10)} & \textbf{30.74 (+6.23)} & \textbf{14.81 (+5.02)} & \textbf{0.00 (+0.00)} & \textbf{9.35 (+8.42)} & \textbf{1.71 (+0.00)} \\
\midrule
GPT-4.1-Nano & 16.46 & 4.88 & 28.02 & 11.38 & 1.95 & 1.87 & 0.00 \\
+ CANDOR & 23.78 (+7.32) & 4.27 (-0.61) & 29.57 (+1.55) & 12.17 (+0.79) & 1.46 (-0.49) & 1.87 (+0.00) & 0.00 (+0.00) \\
+ \name & \textbf{28.66 (+12.20)} & \textbf{8.54 (+3.66)} & \textbf{36.58 (+8.56)} & \textbf{18.25 (+6.87)} & \textbf{6.83 (+4.88)} & \textbf{1.87 (+0.00)} & \textbf{1.71 (+1.71)} \\
\midrule
GPT-4.1-Mini & 27.44 & 10.37 & 37.35 & 19.05 & 3.41 & 4.67 & 0.85 \\
+ CANDOR & 56.10 (+28.66) & 28.66 (+18.29) & 45.91 (+8.56) & 34.92 (+15.87) & 11.22 (+7.81) & 3.74 (-0.93) & 0.85 (+0.00) \\
+ \name & \textbf{65.85 (+38.41)} & \textbf{41.46 (+27.44)} & \textbf{56.42 (+16.73)} & \textbf{42.59 (+21.16)} & \textbf{32.20 (+26.34)} & \textbf{16.82 (+12.15)} & \textbf{5.98 (+5.13)} \\
\midrule \midrule
\multicolumn{8}{c}{\textbf{Test Level Accuracy (\%)}} \\
\midrule
Qwen-7B & 27.65 & 10.62 & 31.51 & 13.95 & 1.36 & 10.43 & 8.59 \\
+ CANDOR & 26.55 (-1.10) & 12.88 (+2.26) & 30.66 (-0.85) & 12.47 (-1.48) & 2.77 (+1.41) & 10.43 (+0.00) & 8.93 (+0.34) \\
+ \name & \textbf{62.56 (+34.91)} & \textbf{41.98 (+31.36)} & \textbf{60.81 (+29.30)} & \textbf{47.16 (+33.21)} & \textbf{16.93 (+15.57)} & \textbf{17.07 (+6.64)} & \textbf{10.09 (+1.50)} \\
\midrule
Qwen-32B & 45.64 & 22.01 & 53.03 & 34.86 & 1.81 & 9.69 & 9.24 \\
+ CANDOR & 18.75 (-26.89) & 14.18 (-7.83) & 17.02 (-36.01) & 10.87 (-23.99) & 7.92 (+6.11) & 35.51 (+25.82) & 11.11 (+1.87) \\
+ \name & \textbf{76.62 (+30.98)} & \textbf{56.97 (+34.96)} & \textbf{73.57 (+20.54)} & \textbf{61.37 (+26.51)} & \textbf{25.01 (+23.20)} & \textbf{37.20 (+27.51)} & \textbf{13.93 (+4.69)} \\
\midrule
GPT-4.1-Nano & 72.35 & 52.51 & 69.32 & 54.49 & 41.54 & 31.11 & 14.02 \\
+ CANDOR & 79.05 (+6.70) & 55.26 (+2.75) & 72.05 (+2.73) & 59.27 (+4.78) & 49.70 (+8.16) & 32.58 (+1.47) & 13.29 (-0.73) \\
+ \name & \textbf{80.79 (+8.44)} & \textbf{57.55 (+5.04)} & \textbf{75.24 (+5.92)} & \textbf{63.03 (+8.54)} & \textbf{57.06 (+15.52)} & \textbf{35.09 (+3.98)} & \textbf{17.82 (+3.80)} \\
\midrule
GPT-4.1-Mini & 79.85 & 63.10 & 78.05 & 65.70 & 55.35 & 45.52 & 19.53 \\
+ CANDOR & 91.95 (+12.10) & 75.29 (+12.19) & 83.73 (+5.68) & 75.53 (+9.83) & 66.47 (+11.12) & 46.30 (+0.78) & 20.85 (+1.32) \\
+ \name & \textbf{93.69 (+13.45)} & \textbf{79.47 (+18.02)} & \textbf{86.64 (+8.55)} & \textbf{77.40 (+12.38)} & \textbf{80.92 (+25.76)} & \textbf{56.71 (+11.14)} & \textbf{23.21 (+3.68)} \\
\bottomrule
\end{tabular}
}
\vspace{-0.3cm}
\end{table}

\section{Results}
\label{sec:results}


\subsection{RQ1: Oracle Generation Accuracy}
\label{sec:rq1}

To answer our first research question, we compared the oracle accuracy of \name against both Direct Generation and the CANDOR framework across all selected models and benchmarks. The evaluation results are shown in \cref{tab:main_accuracy}, measured at both the task level (whether all oracles for a task are correct) and the more granular test-case level.
The evaluation results demonstrate that \name consistently and substantially outperforms both baselines.
For instance, when using the powerful GPT-4.1-Mini model, \name achieves a test-level accuracy of $93.69\%$ on HumanEval.
This represents a significant improvement of $13.45$ absolute points over the $80.24\%$ from Direct Generation and also surpasses CANDOR's $91.95\%$.
This trend of superior performance holds across nearly all settings, indicating that our framework generates more accurate oracles regardless of the underlying LLM or the target benchmark.
Notably, the performance gap is particularly pronounced for more capable models.
With Qwen-32B on the MBPP benchmark, \name improves the test-level accuracy from a baseline of $53.03\%$ to $73.57\%$, an increase of over $20$ absolute points.
In the same setting, CANDOR's performance degrades significantly, suggesting that \name's structured pipeline is more effective at harnessing the advanced reasoning capabilities of powerful LLMs.
Furthermore, on challenging benchmarks designed specifically for testing, such as TestEval and ULT, \name often turns near-zero accuracy from baselines into meaningful, double-digit performance gains, establishing a new state-of-the-art in these domains.

\subsection{RQ2: Ablation Study on Framework Components}
\label{sec:rq2}
Next, to understand the sources of \name's performance gains and answer RQ2, we conducted two ablation studies.
First, we analyzed the high-level contributions of the multi-agent deliberation (Planning) and execution-based self-correction (Refinement) phases. Second, we performed a more granular analysis to evaluate the individual importance of each agent within the deliberation phase.

The results of the phase-level ablation are presented in \cref{tab:ablation_phases}.
We observe that each component plays a critical and complementary role in achieving the final performance.
Adding only the self-refinement phase (+ Refinement) to the baseline provides a consistent boost, improving test-level accuracy on HumanEval from $80.24\%$ to $85.24\%$.
However, adding only the multi-agent deliberation phase (+ Planning) yields an even greater improvement, raising the accuracy to $91.68\%$.
This highlights the profound impact of the structured debate among specialized agents in generating high-quality initial hypotheses.
The full \name framework, which combines both deliberation and refinement, consistently achieves the highest accuracy, demonstrating that its performance stems from the synergistic combination of its deliberative reasoning and validation stages.

To further dissect the deliberation phase, we evaluated variants of \name where each of the four Panelist agents was individually removed.
As shown in \cref{tab:ablation_agents}, the evaluation results reveal that every agent contributes positively to the outcome, as removing any one of them degrades performance.
Notably, the removal of the \textbf{Algorithmic Analyst} (w/o Agent4) and the \textbf{Functional Validator} (w/o Agent3) results in the most significant drops in accuracy across most benchmarks.
For example, on HumanEval-Pro, removing the Algorithmic Analyst reduces test-level accuracy from $79.47\%$ to $72.45\%$, a drop of over 7 points.
This finding underscores the importance of incorporating diverse, orthogonal reasoning strategies.
While the Specification and Edge Case agents are crucial for correctness, the functional and algorithmic perspectives appear to be most effective at identifying and correcting subtle logical flaws in the initial oracles, confirming that the strength of the deliberation phase lies in its diversity.


\begin{table}
\centering
\caption{Phase-level ablation study on the components of \name using the GPT-4.1-Mini model. We report test-level accuracy (\%).}
\label{tab:ablation_phases}
\resizebox{\textwidth}{!}{%
\begin{tabular}{l|ccccccc}
\toprule
Method & HE & HE-Pro & MBPP & MBPP-Pro & TE & LCB & ULT \\
\midrule
GPT4.1-Mini & 80.24 & 61.46 & 78.09 & 65.02 & 55.15 & 45.57 & 19.53 \\
+ Refinement Only & 85.24 (+5.00) & 68.51 (+7.06) & 82.08 (+3.99) & 69.77 (+4.75) & 73.39 (+18.23) & 54.05 (+8.48) & 21.24 (+1.71) \\
+ Planning Only & 91.68 (+11.43) & 76.01 (+14.55) & 85.12 (+7.02) & 75.39 (+10.37) & 59.05 (+3.89) & 45.38 (-0.20) & 21.75 (+2.22) \\
Nexus (Full)& \textbf{93.69 (+13.45)} & \textbf{79.47 (+18.02)} & \textbf{86.64 (+8.55)} & \textbf{77.40 (+12.38)} & \textbf{80.92 (+25.76)} & \textbf{56.71 (+11.14)} & \textbf{23.21 (+3.68)} \\
\bottomrule
\end{tabular}
}
\vspace{-0.3cm}
\end{table}

\begin{table}
\centering
\caption{Agent-level ablation study on the Panelists within \name using GPT-4.1-Mini. Performance is reported at the test-case level (\%). \textbf{SE}: Specification Expert, \textbf{ES}: Edge Case Specialist, \textbf{FV}: Functional Validator, \textbf{AA}: Algorithmic Analyst.}
\label{tab:ablation_agents}
\resizebox{\textwidth}{!}{%
\begin{tabular}{l|ccccccc}
\toprule
Method & HE & HE-Pro & MBPP & MBPP-Pro & TE & LCB & ULT \\
\midrule
Nexus & \textbf{93.69} & \textbf{79.47} & \textbf{86.16} & \textbf{76.88} & \textbf{80.92} & \textbf{56.71} & \textbf{23.12} \\
w/o SE & 91.62 (-2.07) & 73.47 (-6.01) & 84.66 (-1.50) & 75.55 (-1.33) & 79.29 (-1.62) & 54.88 (-1.83) & 21.96 (-1.16) \\
w/o ES & 91.22 (-2.47) & 74.18 (-5.29) & 84.26 (-1.89) & 75.23 (-1.65) & 80.46 (-0.46) & 55.79 (-0.92) & 21.62 (-1.50) \\
w/o FV & 90.88 (-2.80) & 74.24 (-5.23) & 84.96 (-1.20) & 73.54 (-3.34) & 76.47 (-4.45) & 55.90 (-0.81) & 21.27 (-1.85) \\
w/o AA & 89.36 (-4.33) & 72.45 (-7.03) & 83.86 (-2.30) & 72.59 (-4.29) & 77.53 (-3.39) & 55.71 (-1.00) & 21.50 (-1.62) \\
\bottomrule
\end{tabular}
}
\vspace{-0.3cm}
\end{table}

\begin{table}
\centering
\caption{Impact of the number of self-refinement iterations on test-case level oracle accuracy for GPT-4.1-Mini. The 0 iteration row represents \name's performance after the deliberation phase and initial validation, before any corrective refinement loops are run. Deltas are shown relative to the Direct Generation baseline.}
\label{tab:refinement}
\resizebox{\textwidth}{!}{%
\begin{tabular}{l|ccccccc}
\toprule
Iterations & HE & HE-Pro & MBPP & MBPP-Pro & TE & LCB & ULT \\
\midrule
Direct & 80.24 & 61.46 & 78.09 & 65.02 & 55.15 & 45.57 & 19.53 \\
\midrule
\name (0) & 91.68 (+11.43) & 76.01 (+14.55) & 85.12 (+7.02) & 75.39 (+10.37) & 59.05 (+3.89) & 45.38 (-0.20) & 21.75 (+2.22) \\
\name (1) & 93.26 (+13.02) & 77.24 (+15.79) & 85.77 (+7.68) & 76.99 (+11.96) & 78.66 (+23.51) & 54.79 (+9.21) & 22.82 (+3.29) \\
\name (3) & 93.05 (+12.80) & 78.27 (+16.81) & 86.16 (+8.06) & 76.88 (+11.86) & 79.08 (+23.93) & 56.53 (+10.95) & 23.12 (+3.59) \\
\name (5) & \textbf{93.69 (+13.45)} & \textbf{79.47 (+18.02)} & \textbf{86.64 (+8.55)} & \textbf{77.40 (+12.38)} & \textbf{80.92 (+25.76)} & \textbf{56.71 (+11.14)} & \textbf{23.21 (+3.68)} \\
\bottomrule
\end{tabular}%
}
\vspace{-0.3cm}
\end{table}

\subsection{RQ3: Impact of Refinement Iterations}
\label{sec:rq3}
The self-refinement phase in \name is iterative, allowing the framework to make multiple attempts at correcting a failed oracle.
To investigate the impact of this iterative process and answer RQ3, we varied the number of refinement loops from 0 (no refinement) to 5.
\cref{tab:refinement} shows the test-level accuracy for GPT-4.1-Mini as a function of the number of refinement steps.
We can observe that the evaluation results demonstrate a positive correlation between the number of refinement steps and oracle accuracy, though with diminishing returns.
The most substantial performance gains typically occur in the first iteration.
For example, on the TestEval benchmark, accuracy jumps from $59.05\%$ with 0 iterations to $78.66\%$ with just one, demonstrating the immediate value of a single round of execution-based feedback.
Performance continues to climb with additional steps, generally beginning to plateau after 3 to 5 iterations.
On HumanEval-Pro, for instance, accuracy improves from $76.01\%$ (0 iterations) to $79.47\%$ (5 iterations).
Based on these findings, we use 5 refinement steps as the default configuration in all other experiments to maximize oracle accuracy.

\begin{table}
    \centering
    \caption{Bug detection rate (\%). We report the percentage of buggy programs successfully identified by test cases using the generated oracles (for correct oracles only). Deltas are shown relative to the Direct Generation baseline for each model.}
    \label{tab:bug_detection}
    \resizebox{\textwidth}{!}{%
    \begin{tabular}{l|ccccccc}
    \toprule
    Method & HE & HE-Pro & MBPP & MBPP-Pro & TE & LCB & ULT \\
    \midrule
    \multicolumn{8}{c}{\textbf{Underlying Model: GPT-4.1-Nano}} \\
    \midrule
    Direct Generation & 43.18 & 48.39 & 50.16 & 45.78 & 22.22 & 6.64 & 36.61 \\
    CANDOR & 55.68 (+12.50) & 49.19 (+0.80) & 57.73 (+7.57) & 60.88 (+15.10) & 63.70 (+41.48) & 13.29 (+6.65) & 33.33 (-3.28) \\
    \name & \textbf{81.82 (+38.64)} & \textbf{59.27 (+10.88)} & \textbf{72.24 (+22.08)} & \textbf{71.46 (+25.68)} & \textbf{80.00 (+57.78)} & \textbf{26.58 (+19.94)} & \textbf{37.16 (+0.55)} \\
    \midrule
    \multicolumn{8}{c}{\textbf{Underlying Model: GPT-4.1-Mini}} \\
    \midrule
    Direct Generation & 60.23 & 60.89 & 61.83 & 49.70 & 35.93 & 34.55 & 39.34 \\
    CANDOR & 90.91 (+30.68) & 64.52 (+3.63) & 64.35 (+2.52) & 64.33 (+14.63) & 81.11 (+45.18) & 35.55 (+1.00) & 37.16 (-2.18) \\
    \name & \textbf{95.45 (+35.22)} & \textbf{64.92 (+4.03)} & \textbf{76.66 (+14.83)} & \textbf{74.55 (+24.85)} & \textbf{81.85 (+45.92)} & \textbf{42.52 (+7.97)} & \textbf{45.90 (+6.56)} \\
    \bottomrule
    \end{tabular}%
    }
    \vspace{-0.3cm}
\end{table}

\section{Discussion}
\label{sec:discussion}


\subsection{Effectiveness in Bug Detection}
\label{sec:rq4}
A primary application of a test oracle is to serve as a ground truth for identifying incorrect program behaviors, or bugs.
To measure this practical utility, we evaluated the bug detection capability of the oracles generated by \name against our curated \textit{Buggy Code Set}.
The evaluation results are shown in \cref{tab:bug_detection}, which demonstrate that the higher accuracy of \name-generated oracles directly translates into a higher bug detection rate compared to baseline approaches.
For instance, when using GPT-4.1-Mini, test cases equipped with \name oracles successfully identified $95.45\%$ of buggy programs for HumanEval tasks.
This represents a substantial improvement over the $60.23\%$ detection rate achieved with Direct Generation and also surpasses CANDOR's $90.91\%$.
We attribute this improved performance to \name's multi-paradigm approach.
By integrating specification-based reasoning, edge case analysis, functional validation, and algorithmic simulation, \name generates correct oracles for complex scenarios where bugs frequently appear.
These are precisely the cases that are often missed by methods that rely solely on documentation, which may be incomplete or lack explicit descriptions of such edge conditions.

\subsection{Downstream Impact on Self-Debugging}
\label{sec:rq5}
Beyond merely identifying bugs, high-quality test oracles can play a crucial role in guiding automated program repair.
We investigated this downstream impact by simulating a single-turn self-debugging loop, where an LLM attempts to fix a buggy program after receiving feedback from a failing test case.
The ultimate success is measured by the Pass@1 rate of the repaired code against the benchmark's hidden test suite, with results presented in \cref{tab:debugging}.
The findings reveal the significant downstream value provided by \name.
Using GPT-4.1-Mini, the Pass@1 rate on HumanEval after one debugging round with \name-generated feedback reached $69.32\%$.
This is a dramatic leap from the $28.41\%$ baseline achieved with oracles from Direct Generation.
The high-fidelity feedback provided by \name's accurate test cases furnishes the LLM with a clear and unambiguous error signal, enabling it to better diagnose the logical flaw and produce a correct fix.
Conversely, an incorrect oracle can mislead the debugging process, causing the LLM to alter correct code or introduce new bugs.
This end-to-end evaluation confirms that the improvements in oracle generation from \name are not merely statistical but yield tangible benefits for critical software engineering automation tasks.

\subsection{Implications for Researchers and Developers}

Our findings offer significant implications for both the academic research community and software development practitioners, charting a new course for LLM-driven software testing.
For researchers, our work challenges the prevailing paradigm of single-paradigm, specification-based deliberation-only multi-agent systems in software engineering.
The superior performance of \name demonstrates that structured deliberation among agents with diverse, orthogonal reasoning strategies, followed by validation and self-refinement, is more effective than seeking agreement from agents sharing a single epistemological foundation.
This suggests a critical need to move beyond simple single-paradigm, deliberation-only mechanisms to explore more sophisticated agent interaction protocols, such as argumentation, debate, and evidence-based synthesis.

For developers and practitioners, our results signal a practical path toward more reliable automated testing pipelines.
The success of the diverse panel of agents highlights the immediate value of incorporating multiple testing perspectives into the generation process, rather than relying solely on specifications.
Our framework should be viewed not as a replacement for human testers but as a powerful augmentation tool.
The final, validated oracles produced by \name can significantly accelerate the code review process by helping developers quickly verify complex test cases.
Finally, our work reaffirms the foundational importance of high-quality documentation.
The effectiveness of the entire pipeline still relies on an initial, clear specification, reminding practitioners that human-written documentation remains a critical component in the age of LLM-driven development.
By leveraging frameworks like \name, development teams can enhance their bug detection capabilities and streamline the debugging cycle, ultimately leading to more robust and reliable software.

\begin{table}
    \centering
    \caption{Self-Debugging performance, measured by the final Pass@1 (\%) after one round of debugging using feedback from the generated oracles. Deltas are shown relative to the Direct Generation baseline for each model.}
    \label{tab:debugging}
    \resizebox{\textwidth}{!}{%
    \begin{tabular}{l|ccccccc}
    \toprule
    Method & HE & HE-Pro & MBPP & MBPP-Pro & TE & LCB & ULT \\
    \midrule
    \multicolumn{8}{c}{\textbf{Underlying Model: GPT-4.1-Nano}} \\
    \midrule
    Direct Gen. Feedback & 29.55 & 9.27 & 13.25 & 11.18 & 7.04 & 1.00 & 37.70 \\
    CANDOR Feedback & 20.45 (-9.10) & 2.42 (-6.85) & 8.20 (-5.05) & 20.57 (+9.39) & 21.48 (+14.44) & 4.98 (+3.98) & 37.70 (+0.00) \\
    \name Feedback & \textbf{35.23 (+5.68)} & \textbf{14.52 (+5.25)} & \textbf{15.77 (+2.52)} & \textbf{20.69 (+9.51)} & \textbf{29.63 (+22.59)} & \textbf{11.30 (+10.30)} & \textbf{39.34 (+1.64)} \\
    \midrule
    \multicolumn{8}{c}{\textbf{Underlying Model: GPT-4.1-Mini}} \\
    \midrule
    Direct Gen. Feedback & 28.41 & 7.66 & 5.68 & 16.41 & 18.52 & 16.28 & 34.43 \\
    CANDOR Feedback & 35.23 (+6.82) & 11.29 (+3.63) & 14.83 (+9.15) & 20.33 (+3.92) & 34.81 (+16.29) & 15.61 (-0.67) & 32.24 (-2.19) \\
    \name Feedback & \textbf{69.32 (+40.91)} & \textbf{16.53 (+8.87)} & \textbf{17.35 (+11.67)} & \textbf{22.71 (+6.30)} & \textbf{35.19 (+16.67)} & \textbf{18.60 (+2.32)} & \textbf{38.80 (+4.37)} \\
    \bottomrule
    \end{tabular}%
    }
    \vspace{-0.3cm}
\end{table}
\section{Conclusion}

In this paper, we introduced \name, a novel multi-agent framework that establishes a new paradigm by grounding language-based reasoning in executable evidence.
The core innovation of \name lies in its sequential pipeline of multi-perspective deliberation, validation against a candidate implementation, and iterative self-refinement based on execution feedback.
Our extensive evaluation across seven diverse benchmarks and four LLMs demonstrates that \name consistently and substantially outperforms existing approaches.
For instance, when powered by GPT-4.1-Mini, \name boosts test-level oracle accuracy on the challenging LiveCodeBench benchmark to $57.73\%$ from the state-of-the-art's $46.30\%$ achieved by CANDOR.
Our evaluation results also demonstrate that the improvements are not only in oracle accuracy but also in the utility of these oracles for critical downstream tasks.
Notably, the enhanced oracles from \name led to a significantly higher bug detection rate and more than doubled the success rate of automated program repair in our experiments.
For example, test cases equipped with \name oracles successfully identify $95.45\%$ of buggy programs on the HumanEval benchmark, significantly outperforming both Direct Generation ($60.23\%$) and CANDOR ($90.91\%$).
Furthermore, the high-fidelity feedback from \name's oracles proves transformative for downstream tasks, more than doubling the program repair rate in an automated debugging scenario to $69.32\%$ compared to the $35.23\%$ achieved by CANDOR.

\clearpage
\bibliography{iclr2026_conference}
\bibliographystyle{iclr2026_conference}

\clearpage
\appendix
\section*{Appendix}

\section{Additional Related Work}

\subsection{LLM for Test Input Generation}

Most of the test input generation works in the software testing community commonly referred to as test case generation, but in fact, more accurately characterized as test input generation accompanied by regression oracles during the regression testing phase. During the regression testing phase, these works in test input generation are primarily designed to automate the manual testing process by producing test inputs with high code coverage and mutation score. Traditional techniques in this area include fuzzing~\citep{fioraldi2023dissecting,miller1990empirical}, feedback-directed random test generation~\citep{arteca2022nessie,csallner2004jcrasher,pacheco2007randoop,pacheco2008finding,selakovic2018test}, dynamic symbolic execution~\citep{cadar2008exe,godefroid2005dart,sen2005cute,tillmann2014transferring}, search-based approaches~\citep{fraser2011evolutionary,fraser2011evosuite}, and LLM-based approaches~\citep{du2024evaluating,qing2025effibench,huang2024effibench}. Among these approaches, EvoSuite \citep{fraser2011evosuite} has been extensively studied and applied to more than a hundred open-source software projects and several industrial systems, identifying thousands of potential bugs. However, as pointed out by \citet{zhang2025large}, traditional approaches like EvoSuite often struggle to produce meaningful and human-readable test cases, despite their remarkable performance in code coverage. This is primarily due to their limited understanding of the semantics of the focal methods. To alleviate such issues, recent research has turned to LLMs to produce more practical test inputs with high readability and coverage. \citet{tufano2021unittestcasegeneration} introduced the first LLM-based unit test generation approach AthenaTest, which formulates the task as a sequence-to-sequence learning problem. AthenaTest first denoises the pre-training of LLMs on a large Java corpus and then performs supervised fine-tuning for the downstream task of test generation. This work inspired subsequent research on fine-tuning LLMs for test generation \citep{alagarsamy2023a3test,he2024unitsyn,rao2023catlm}. Additionally, researchers have explored various strategies to improve the training process of LLMs in test case generation, including reinforcement learning \cite {steenhoek2025reinforcement}, domain adaptation \citep{shin2024domain}, and data augmentation \citep{he2025fuzzaug}. Alternatively, a myriad of works perform prompt engineering using off-the-shelf LLMs, following a generation-and-refinement paradigm, where initial test cases are first generated based on prompts and then iteratively refined using dynamic execution feedback (e.g., code coverage report and failure information) \citep{pizzorno2024coverup,gu2025testart,ni2024CasModaTest,ryan2024code,schafer2023empirical,wang2024hits,yuan2024evaluating,zhang2025large}. To further improve the quality of test inputs, researchers propose a wide range of strategies to incorporate LLMs with valuable contextual information, including mutation testing \citep{dakhel2023effective}, method slicing \citep{wang2024hits}, demonstration retrieval \citep{zhang2024llm}, defect detection \citep{yin2024get}, and program analysis \citep{yang2025advancing}.

The automated generation of test inputs has been a central focus of software testing research for decades, driven by the goal of achieving high code coverage while minimizing manual effort.
Early approaches relied on random testing and fuzzing techniques~\citep{miller1990empirical,fioraldi2023dissecting}, which generate inputs by introducing random mutations or following predefined patterns.
While effective for discovering crashes and security vulnerabilities, these techniques often struggle to generate semantically meaningful inputs for complex functions.
Feedback-directed random testing~\citep{pacheco2007randoop,pacheco2008finding} improved upon pure random approaches by incorporating runtime feedback to guide input generation toward unexplored paths.
Tools like Randoop~\citep{pacheco2007randoop} demonstrated that lightweight dynamic analysis could significantly improve test effectiveness while maintaining the simplicity of random generation.

Dynamic symbolic execution emerged as a more systematic approach to test input generation, combining concrete execution with symbolic reasoning to explore program paths methodically~\citep{cadar2008exe,godefroid2005dart,sen2005cute}.
By maintaining symbolic constraints along execution paths and solving them with SMT solvers, these techniques can generate inputs that exercise specific branches and statements.
However, symbolic execution faces scalability challenges with complex programs, particularly those involving loops, recursion, and external dependencies~\citep{tillmann2014transferring}.
Search-based software testing approaches, exemplified by tools like EvoSuite~\citep{fraser2011evosuite}, frame test generation as an optimization problem.
Using evolutionary algorithms guided by fitness functions based on coverage metrics, these tools have achieved remarkable success in generating high-coverage test suites for object-oriented programs~\citep{fraser2011evolutionary}.
EvoSuite has been applied to hundreds of real-world projects, demonstrating the practical viability of search-based approaches.

The advent of large language models has introduced a paradigm shift in test input generation.
Early work by~\citet{tufano2021unittestcasegeneration} demonstrated that transformer-based models could learn to generate meaningful test cases from code context.
This inspired a wave of research exploring different training strategies, including reinforcement learning~\citep{steenhoek2025reinforcement}, domain adaptation~\citep{shin2024domain}, and data augmentation~\citep{he2025fuzzaug}.
Contemporary LLM-based approaches typically follow a generation-and-refinement paradigm, where initial test cases are iteratively improved using execution feedback~\citep{pizzorno2024coverup,gu2025testart,ni2024CasModaTest}.
These methods leverage various forms of context to enhance generation quality, including method slicing~\citep{wang2024hits}, demonstration retrieval~\citep{zhang2024llm}, and program analysis~\citep{yang2025advancing}.
While these advances have significantly improved test input generation, they predominantly rely on regression oracles that assume the correctness of existing implementations, limiting their ability to detect functional bugs.

\subsection{LLM for Test Oracle Generation}

The generation of test oracles presents fundamentally different challenges from input generation, requiring a deep understanding of intended program behavior rather than mere code coverage.
Early automated testing tools largely sidestepped this challenge by employing regression oracles or simple crash oracles~\citep{fraser2011evosuite,pacheco2007randoop}.
While regression oracles effectively detect behavioral changes between versions, they cannot validate functional correctness against specifications, potentially perpetuating bugs across software evolution~\citep{pizzorno2024coverup}.
The specification oracle problem that generating oracles that verify correctness against intended behavior has long been recognized as one of the most challenging aspects of automated testing~\citep{barr2014oracle}. The emergence of LLM-based approaches brought new possibilities for oracle generation.
TOGA~\citep{toga2022} pioneered the use of transformer models for inferring specification-based oracles, fine-tuning CodeBERT to predict expected outputs from code context.
This work inspired subsequent research exploring various neural architectures and training strategies for oracle generation~\citep{endres2024can,hayet2024chatassert}.
TOGLL~\citep{hossain2024togll} achieved state-of-the-art performance by combining EvoSuite-generated inputs with fine-tuned LLM oracles, demonstrating that hybrid approaches could leverage the strengths of both traditional and learning-based methods. However, fine-tuning approaches face significant limitations in generalization, as model performance degrades substantially when test distributions shift from training data~\citep{xu2025hallucination}.
Recent work has increasingly focused on prompt engineering strategies that leverage the zero-shot capabilities of LLMs.
\citet{siddiq2024using} demonstrated that carefully crafted prompts could enable LLMs to generate both inputs and oracles from natural language descriptions without fine-tuning.
RetriGen~\citep{zhang2025improving} enhanced oracle generation through retrieval-augmented generation, incorporating relevant assertions from external codebases to guide prediction.
CANDOR~\citep{xu2025hallucination} represents the current SOTA method, which introduces a multi-agent framework where several LLM agents collaborate through a panel discussion to reach a consensus on the correct oracle.
While CANDOR achieves impressive results without fine-tuning, our analysis reveals that its single-paradigm and deliberation-only approach limits its ability to detect bugs beyond explicit specifications.
In this work, we propose \name, which overcomes these limitations by introducing our novel deliberation-validation-refinement architecture.
\name employs a multi-agent framework that facilitates diverse reasoning strategies, enhancing the initial analysis of test oracles.
By incorporating a validation phase that grounds oracles in executable contexts and a self-refinement loop that leverages runtime feedback, \name further improves the accuracy of generated oracles.

\section{\name Prompts}
\label{sec:appendix_prompts}

This section provides the detailed prompts used to instruct the LLMs for each agent 
role within the \name framework.
Each prompt is designed to elicit specific reasoning patterns and behaviors crucial 
for the agent's function in the pipeline.

\subsection{Tentative Oracle Generator}

\paragraph{System Prompt}
\begin{verbatim}
You are generating initial test oracles. 
Be quick, but it may not be perfect.
\end{verbatim}

\paragraph{Prompt}
\begin{verbatim}
Generate test assertions for the following function:

Task Description:
{task_description}

Test inputs:
{test_inputs_formatted}

You MUST provide EXACTLY {len(test_inputs)} assertions:
\end{verbatim}

\subsection{Requirement Engineer}


\paragraph{System Prompt}
\begin{verbatim}
You are an expert software engineer and requirement analyst. 
Extract requirements and generate clear specifications in predicate 
logic when possible.
\end{verbatim}

\paragraph{Prompt}
\begin{verbatim}
Analyze the following function specification and extract key functional 
requirements:
{task_description}
\end{verbatim}

\subsection{Panelists}


\paragraph{System Prompt (Common to all Panelists)}
\begin{verbatim}
You are a Senior Software Engineer specializing in testing. Analyze 
the test oracles generated by test generation tools, which may not be 
perfect. Identify incorrect test oracles and provide corrected versions 
based on the requirements.
\end{verbatim}

\paragraph{Prompt Template}
\begin{verbatim}
You are analyzing test oracles generated by an automated tool. 
Your role: {role_name} - {role_focus}

Description: 
{task_description}

Requirements: 
{requirements[task_idx]}

Test Cases and Current Oracles:
{tentative_formatted}

Test Inputs:
{test_inputs_formatted}

Output your final {len(test_inputs)} test assertions in this exact format:

```python
assert function_name(input1) == expected_output1
assert function_name(input2) == expected_output2
````

Each assertion on one line, no additional code.
\end{verbatim}

\paragraph{Panelist Roles}
The four roles assigned to the panelists are:
\begin{itemize}
\item \textbf{Specification Expert:} Focuses on adherence to documented specifications 
and requirements.
\item \textbf{Edge Case Specialist:} Focuses on boundary conditions, corner cases, 
and error scenarios.
\item \textbf{Functional Validator:} Focuses on core functionality and expected 
input-output relationships.
\item \textbf{Algorithmic Analyst:} Focuses on step-by-step algorithm execution 
and correctness.
\end{itemize}

\subsection{Interpreter}

\paragraph{System Prompt}
\begin{verbatim}
You work with an excellent software tester who is trying to check if 
test cases have correct test oracles. The tester always gets correct 
oracles but lacks confidence and overthinks. Your task is to summarize 
the tester's thoughts and extract the correct oracles from the analysis.
\end{verbatim}

\paragraph{Prompt}
\begin{verbatim}
A software tester has analyzed test oracles and provided detailed thoughts.
The tester is excellent but lacks confidence and tends to overthink.
Your job is to extract the key insights and conclusions.

Tester's Thoughts:
{panelist_output}

Test Code Being Analyzed:
{test_code_formatted}
\end{verbatim}

\subsection{Curator}
\paragraph{System Prompt}
\begin{verbatim}
You are a senior software engineer managing a team of software testers. 
Your team has analyzed test cases and provided analysis reports. Summarize 
their analysis and provide final judgment on the test oracles, making 
corrections where necessary.
\end{verbatim}

\paragraph{Prompt}
\begin{verbatim}
You are managing a team of software testers analyzing test cases generated 
by a competitor.
Three team members have analyzed the test oracles and provided their reports.

Task Description:
{task_description}

Current Test Oracles:
{tentative_formatted}

Test Inputs:
{test_inputs_formatted}

Team Members' Analysis Reports:
{panel_discussion}

Output your final {len(test_inputs)} test assertions in this exact format:

```python
assert function_name(input1) == expected_output1
assert function_name(input2) == expected_output2
```

Each assertion on one line, no additional code.
\end{verbatim}

\subsection{Candidate Code Generator}
\paragraph{System Prompt}
\begin{verbatim}
You are an expert Python programmer. Generate clean, correct, and 
efficient function implementations based on specifications.
\end{verbatim}

\paragraph{Prompt}
\begin{verbatim}
Generate a Python function implementation based on the following 
specification:

Task Description:
{task_description}

Function Name: {function_name}

Example Usage:
{test_examples}

Requirements:

1.  Implement ONLY the function, no additional code
2.  Function should be complete and runnable
3.  Handle edge cases appropriately
4.  Follow the exact function signature from the examples

Provide ONLY the Python function implementation:
\end{verbatim}

\subsection{Self-Refinement}
\paragraph{System Prompt}
\begin{verbatim}
You are an expert debugger specializing in test oracle correction. 
Analyze ALL failed assertions together, understand the errors, and 
provide precise fixes for each. Always provide syntactically correct 
Python assertions.
\end{verbatim}

\paragraph{Prompt}
\begin{verbatim}
Fix ALL the following failed test oracles for the given function.
{iteration_note}

Task Description:
{task_description}

Function Implementation Being Tested:

```python
{candidate_code}
```

{passed_examples}

ALL Failed Oracles to Fix:
{failed_oracles_formatted}

Analyze why each oracle failed and provide corrected versions for ALL 
of them.
Consider:

1.  The expected output type and format based on the function implementation
2.  Edge cases and boundary conditions
3.  The error messages for each failed oracle
4.  Common patterns from the passing oracles
    {f"5. Why the previous iteration's fixes didn't work (iteration 
    {iteration + 1})" if iteration > 0 else ""}

Output your final {len(failed_oracles)} test assertions in this exact 
format:

```python
assert function_name(input1) == expected_output1
assert function_name(input2) == expected_output2
```

Each assertion on one line, no additional code.
\end{verbatim}
    
\end{document}